\documentclass[vecphys]{svmult}

\usepackage{graphicx}        
\usepackage{multicol}        

\makeindex             

\begin{document}

\title*{Field-theoretic Methods}

\author{Uwe Claus T\"auber}

\institute{Department of Physics,
        Center for Stochastic Processes in Science and Engineering \\ 
        Virginia Polytechnic Institute and State University \\
        Blacksburg, Virginia 24061-0435, USA \\
	email: \texttt{tauber@vt.edu}}

\begin{abstract}
Many complex systems are characterized by intriguing spatio-temporal 
structures. Their mathematical description relies on the analysis of 
appropriate correlation functions. Functional integral techniques provide a 
unifying formalism that facilitates the computation of such correlation 
functions and moments, and furthermore allows a systematic development of 
perturbation expansions and other useful approximative schemes. It is 
explained how nonlinear stochastic processes may be mapped onto exponential 
probability distributions, whose weights are determined by continuum field 
theory actions. Such mappings are madeexplicit for (1) stochastic interacting 
particle systems whose kinetics is defined through a microscopic master 
equation; and (2) nonlinear Langevin stochastic differential equations which 
provide a mesoscopic description wherein a separation of time scales between 
the relevant degrees of freedom and background statistical noise is assumed. 
Several well-studied examples are introduced to illustrate the general 
methodology.
\end{abstract}

\maketitle

\phantom{} \smallskip
\section*{Article Outline}
\phantom{} \ Glossary \smallskip
\begin{enumerate}
\item Fluctuations and correlations, field-theoretic methods \\
\item Introduction \\
      2.1. Stochastic complex systems \\
      2.2. Example: Lotka--Volterra model \\
\item Correlation functions and field theory \\
      3.1. Generating functions \\
      3.2. Perturbation expansion \\
      3.3. Continuum limit and functional integrals \\
\item Discrete stochastic interacting particle systems \\ 
      4.1. Master equation and Fock space representation \\
      4.2. Continuum limit and field theory \\
      4.3. Annihilation processes \\
      4.4. Active to absorbing state phase transitions \\
\item Stochastic differential equations \\ 
      5.1. Field theory representation of Langevin equations \\
      5.2. Thermal equilibrium and relaxational critical dynamics \\
      5.3. Driven diffusive systems and interface growth \\
\item Future Directions \\ 
\item Bibliography
\end{enumerate}

\section*{Glossary}

\begin{itemize}

\item[] {\bf Absorbing state} \\
	State from which, once reached, an interacting many-particle system 
	cannot depart, not even through the aid of stochastic fluctuations. 
\smallskip

\item[] {\bf Correlation function} \\
      	Quantitative measures of the correlation of random variables; usually 
      	set to vanish for statistically independent variables.
\smallskip

\item[] {\bf Critical dimension} \\
        Borderline dimension $d_c$ above which mean-field theory yields 
	reliable results, while for $d \leq d_c$ fluctuations crucially affect 
	the system's large scale behavior.
\smallskip

\item[] {\bf External noise} \\
	Stochastic forcing of a macroscopic system induced by random external 
	perturbations, such as thermal noise from a coupling to a heat bath.
\smallskip

\item[] {\bf Field theory} \\
	A representation of physical processes through continuous variables, 
  	typically governed by an exponential probability distribution.
\smallskip

\item[] {\bf Generating function} \\
	Laplace transform of the probability distribution; all moments and 
	correlation functions follow through appropriate partial derivatives.
\smallskip

\item[] {\bf Internal noise} \\
	Random fluctuations in a stochastic macroscopic system originating 
	from its internal kinetics.
\smallskip

\item[] {\bf Langevin equation} \\
	Stochastic differential equation describing time evolution that is
	subject to fast random forcing.
\smallskip

\item[] {\bf Master equation} \\
	Evolution equation for a configurational probability obtained by 
	balancing gain and loss terms through transitions into and away from 
	each state.
\smallskip

\item[] {\bf Mean-field approximation} \\
	Approximative analytical approach to an interacting system with many 
	degrees of freedom wherein spatial and temporal fluctuations as well as
	correlations between the constituents are neglected. 
\smallskip

\item[] {\bf Order parameter} \\
	A macroscopic density corresponding to an extensive variable that 
	captures the symmetry and thereby characterizes the ordered state of a 
	thermodynamic phase in thermal equilibrium. 
	Nonequilibrium generalizations typically address appropriate stationary
	values in the long-time limit.
\smallskip

\item[] {\bf Perturbation expansion} \\
	Systematic approximation scheme for an interacting and / or nonlinear 
	system that involves a formal expansion about an exactly solvable 
	simplication by means of a power series with respect to a small 
        coupling.

\end{itemize}

\section{Fluctuations and correlations, field-theoretic methods}
\label{sec:flcor}

Traditionally, complex macroscopic systems are often described in terms of 
ordinary differential equations for the temporal evolution of the relevant 
(usually collective) variables.
Some natural examples are particle or population densities, chemical reactant 
concentrations, and magnetization or polarization densities; others involve
more abstract concepts such as an apt measure of activity, etc.
Complex behavior often entails (diffusive) spreading, front propagation, and 
spontaneous or induced pattern formation.
In order to capture these intriguing phenomena, a more detailed level of 
description is required, namely the inclusion of spatial degrees of freedom, 
whereupon the above quantities all become local density fields.
Stochasticity, i.e., randomly occuring propagation, interactions, or reactions,
frequently represents another important feature of complex systems.
Such stochastic processes generate {\em internal noise} that may crucially 
affect even long-time and large-scale properties.
In addition, other system variables, provided they fluctuate on time scales 
that are fast compared to the characteristic evolution times for the relevant 
quantities of interest, can be (approximately) accounted for within a Langevin 
description in the form of {\em external} additive of multiplicative noise.

A quantitative mathematical analysis of complex spatio-temporal structures and
more generally cooperative behavior in stochastic interacting systems with many
degrees of freedom typically relies on the study of appropriate 
{\em correlation functions}.
{\em Field-theoretic}, i.e., spatially continuous, representations both for 
random processes defined through a master equation and Langevin-type stochastic
differential equations have been developed since the 1970s.
They provide a general framework for the computation of correlation functions,
utilizing powerful tools that were originally developed in quantum many-body 
as well as quantum and statistical field theory. 
These methods allow us to construct systematic approximation schemes, e.g.,  
{\em perturbative expansions} with respect to some parameter (presumed small) 
that measures the strength of fluctuations.
They also form the basis of more sophisticated renormalization group methods 
which represent an especially potent device to investigate scale-invariant 
phenomena.

\section{Introduction}
\label{sec:intro}

\subsection{Stochastic complex systems}
\label{sus:stocs}

Complex systems consist of many interacting components.
As a consequence of either these interactions and / or the kinetics governing
the system's temporal evolution, correlations between the constituents emerge
that may induce cooperative phenomena such as (quasi-)periodic oscillations,
the formation of spatio-temporal patterns, and phase transitions between 
different macroscopic states.
These are characterized in terms of some appropriate collective variables, 
often termed {\em order parameters}, which describe the large-scale and 
long-time system properties.
The time evolution of complex systems typically entails random components:
either, the kinetics itself follows stochastic rules (certain processes occur 
with given probabilities per unit time); or, we project our ignorance of 
various fast microscopic degrees of freedom (or our lack of interest in their 
detailed dynamics) into their treatment as stochastic noise.

An exact mathematical analysis of nonlinear stochastic systems with many 
interacting degrees of freedom is usually not feasible.
One therefore has to resort to either computer simulations of corresponding
stochastic cellular automata, or approximative treatments.
A first step, which is widely used and often provides useful qualitative
insights, consists of ignoring spatial and temporal fluctuations, and just
studying equations of motion for ensemble-averaged order parameters.
In order to arrive at closed equations, additional simplifications tend to be
necessary, namely the factorization of correlations into powers of the mean
order parameter densities.
Such approximations are called {\em mean-field} theories; familiar examples are
rate equations for chemical reaction kinetics or Landau--Ginzburg theory for 
phase transitions in thermal equilibrium.
Yet in some situations mean-field approximations are insufficient to obtain a 
satisfactory quantitative description (see, e.g., the recent work collected in 
Refs.~\cite{JCM, JSP}).
Let us consider an illuminating example.

\subsection{Example: Lotka--Volterra model}
\label{sus:lvmod}

In the 1920s, Lotka and Volterra independently formulated a mathematical model
to describe emerging periodic oscillations respectively in coupled 
autocatalytic chemical reactions, and in the Adriatic fish population (see, 
e.g., Murray 2002 \cite{Murray}).
We shall formulate the model in the language of population dynamics, and treat
it as a stochastic system with two species $A$ (the `predators') and $B$ (the
`prey'), subject to the following reactions: predator death $A \to \emptyset$,
with rate $\mu$; prey proliferation $B \to B + B$, with rate $\sigma$; 
predation interaction $A + B \to A + A$, with rate $\lambda$.
Obviously, for $\lambda = 0$ the two populations decouple; while the predators 
face extinction, the prey population will explode. 
The average predator and prey population densities $a(t)$ and $b(t)$ are 
governed by the linear differential equations $\dot{a}(t) = - \mu \, a(t)$
and $\dot{b}(t) = \sigma \, b(t)$, whose solutions are exponentials.
Interesting competition arises as a consequence of the nonlinear process 
governed by the rate $\lambda$.
In an exact representation of the system's temporal evolution, we would now 
need to know the probability of finding an $A$-$B$ pair at time $t$.
Moreover, in a spatial Lotka--Volterra model, defined on a $d$-dimensional
lattice, say, on which the individual particles can move via nearest-neighbor
hopping, the predation reaction should occur only if both predators and 
prey occupy the same or adjacent sites.
The evolution equations for the mean densities $a(t)$ and $b(t)$ would then 
have to be respectively amended by the terms $\pm \lambda \langle a(x,t) \, 
b(x,t) \rangle$.
Here $a(x,t)$ and $b(x,t)$ represent local concentrations, the brackets denote 
the ensemble average, and $\langle a(x,t) \, b(x,t) \rangle$ represents $A$-$B$
cross correlations.

In the rate equation approximation, it is assumed that the local densities are
uncorrelated, whereupon $\langle a(x,t) \, b(x,t) \rangle$ factorizes to 
$\langle a(x,t) \rangle \, \langle b(x,t) \rangle = a(t) \, b(t)$.
This yields the famous deterministic Lotka--Volterra equations
\begin{equation}
  \dot{a}(t) = \lambda \, a(t) \, b(t) - \mu \, a(t) \ , \quad 
  \dot{b}(t) = \sigma \, b(t) - \lambda \, a(t) \, b(t) \ .
\label{lotvol}
\end{equation}
Within this mean-field approximation, the quantity 
$K(t) = \lambda [a(t) + b(t)] - \sigma \ln{a(t)} - \mu \ln{b(t)}$ 
(essentially the system's Lyapunov function) is a constant of motion, 
$\dot{K}(t) = 0$.
This results in regular nonlinear population oscillations, whose frequency and 
amplitude are fully determined by the initial conditions, a rather unrealistic 
feature. 
Moreover Eqs.~(\ref{lotvol}) are known to be unstable with respect to various
model modifications (as discussed in Murray 2002 \cite{Murray}).

\begin{figure} \centering 
\includegraphics[width = 11.8 truecm]{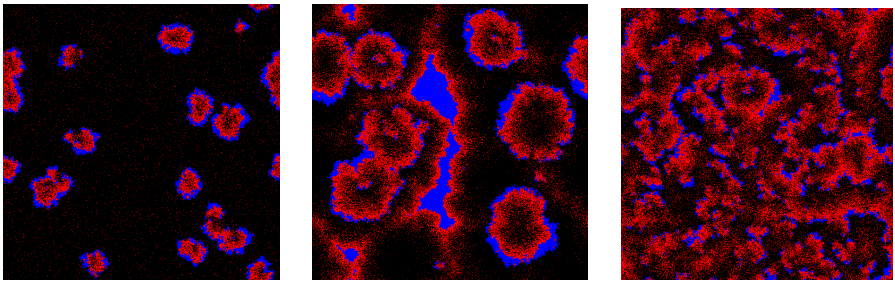} 
\caption{{\bf Field-theoretic Methods.}
        Snapshots of the time evolution (left to right) of activity fronts 
	emerging in a stochastic Lotka--Volterra model simulated on a 
	$512 \times 512$ lattice, with periodic boundary conditions and site 
	occupation numbers restricted to $0$ or $1$. For the chosen reaction
	rates, the system is in the species coexistence phase (with rates
	$\sigma = 4.0$, $\mu = 0.1$, and $\lambda = 2.2$), and the 
	corresponding mean-field fixed point a focus. The red, blue, and black
	dots respectively represent predators $A$, prey $B$, and empty sites 
	$\emptyset$. Reproduced with permission from Ref.~\cite{MobGeoTau}.} 
\label{2dsnap} 
\end{figure} 

In contrast with the rate equation predictions, the original stochastic spatial
Lotka--Volterra system displays much richer behavior (a recent overview is
presented in Ref.~\cite{MobGeoTau}):
The predator--prey coexistence phase is governed, for sufficiently large values
of the predation rate, by an incessant sequence of `pursuit and evasion' wave
fronts that form quite complex dynamical patterns, as depicted in 
Figure~\ref{2dsnap}, which shows snapshots taken in a two-dimensional lattice
Monte Carlo simulation where each site could at most be occupied by a single
particle.
In finite systems, these correlated structures induce erratic population 
oscillations whose features are independent of the initial configuration.
Moreover, if locally the prey `carrying capacity' is limited (corresponding to
restricting the maximum site occupation number per lattice site), there appears
an extinction threshold for the predator population that separates the active 
coexistence regime through a continuous phase transition from a state wherein 
at long times $t \to \infty$ only prey survive.
With respect to the predator population, this represents an {\em absorbing 
state}: once all $A$ particles have vanished, they cannot be produced by the 
stochastic kinetics.

\begin{figure} \centering
\includegraphics[width = 5.5cm]{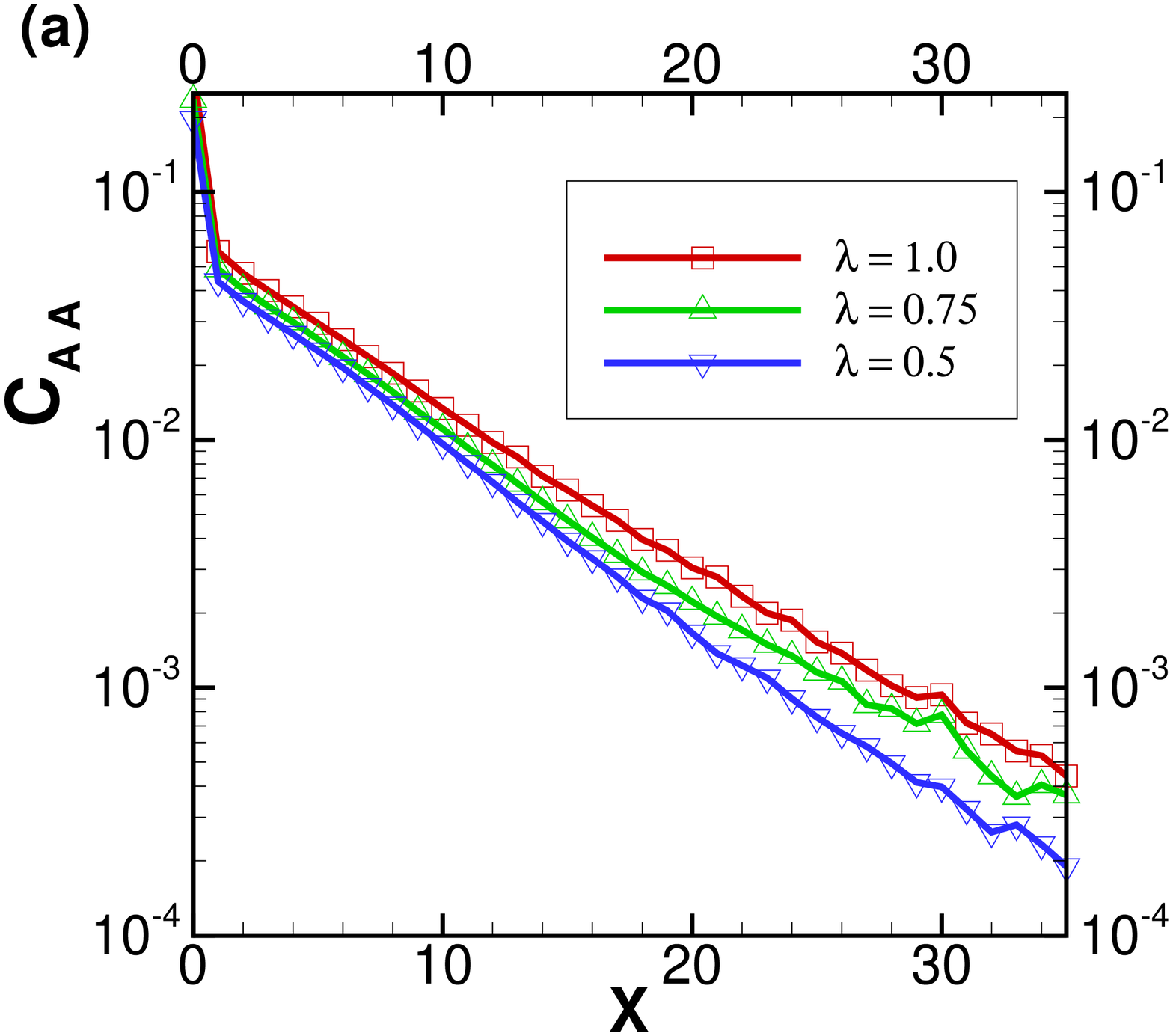} \quad
\includegraphics[width = 5.5cm]{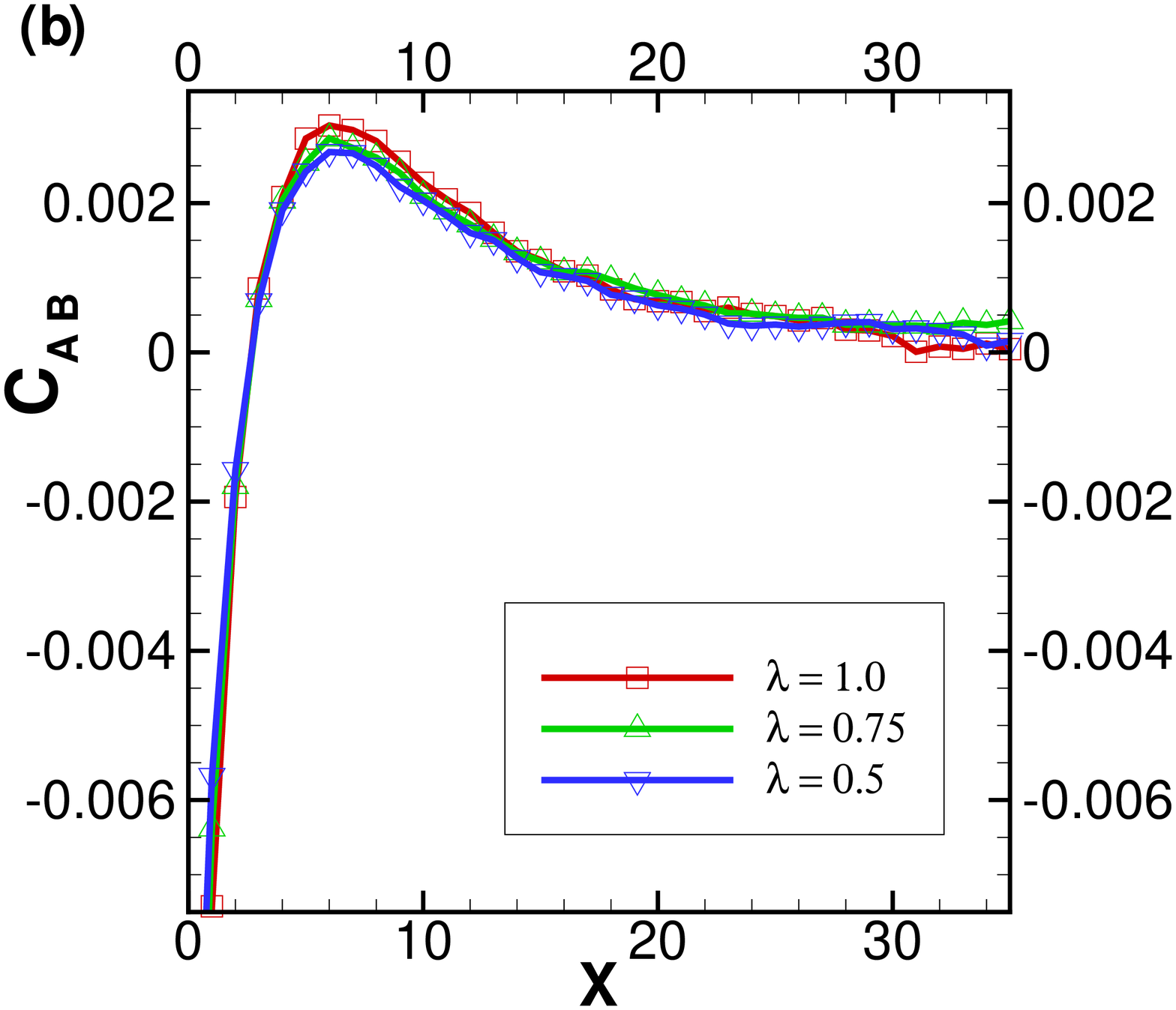}
\caption{{\bf Field-theoretic Methods.}
        Static correlation functions (a) $C_{AA}(x)$ (note the logarithmic 
        scale), and (b) $C_{AB}(x)$, measured in simulations on a 
	$1024 \times 1024$ lattice without any restrictions on the site 
	occupations. The reaction rates were $\sigma = 0.1$, $\mu = 0.1$, and 
	$\lambda$ was varied from $0.5$ (blue triangles, upside down), $0.75$ 
	(green triangles), to $1.0$ (red squares).
	Reproduced with permission from Ref.~\cite{WasMobTau}.}
\label{abcorr}
\end{figure}

\begin{figure} 
\centering 
\includegraphics[width = 4.5 truecm]{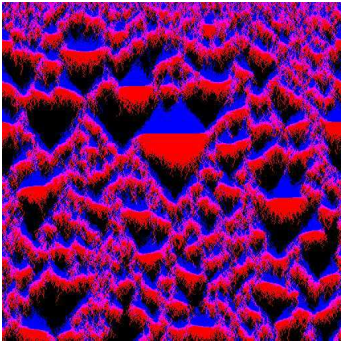} 
\caption{{\bf Field-theoretic Methods.}
        Space-time plot (space horizontal, with periodic boundary conditions;
        time vertical, proceeding downward) showing the temporal evolution of a
        one-dimensional stochastic Lotka--Volterra model on $512$ lattice 
	sites, but without any restrictions on the site occupation numbers 
	(red: predators, blue: prey, magenta: sites occupied by both species; 
	rates: $\sigma = 0.1$, $\mu = 0.1$, $\lambda = 0.1$). 
	Reproduced with permission from Ref.~\cite{WasMobTau}.} 
\label{1dabrc} 
\end{figure}

A quantitative characterization of the emerging spatial structures utilizes
equal-time correlation functions such as $C_{AA}(x-x',t) = \langle a(x,t) \, 
a(x',t) \rangle - a(t)^2$ and $C_{AB}(x-x',t) = \langle a(x,t) \, b(x',t) 
\rangle - a(t) \, b(t)$, computed at some large time $t$ in the 
(quasi-)stationary state.
These are shown in Figure~\ref{abcorr} as measured in computer simulations for 
a stochastic Lotka--Volterra model (but here no restrictions on the site
occupation numbers of the $A$ or $B$ particles were implemented).
The $A$-$A$ (and $B$-$B$) correlations obviously decay essentially 
exponentially with distance $x$, $C_{AA}(x) \propto C_{BB}(x) \propto 
e^{-|x|/\xi}$, with roughly equal correlation lengths $\xi$ for the predators 
and prey.
The cross-correlation function $C_{AB}(x)$ displays a maximum at six lattice 
spacings; these positive correlations indicate the spatial extent of the 
emerging activity fronts (prey followed by the predators).
At closer distance, the $A$ and $B$ particles become {\em anti}-correlated 
($C_{AB}(x) < 0 $ for $|x| < 3$): prey would not survive close encounters with
the predators.
In a similar manner, one can address temporal correlations.
These appear prominently in the space-time plot of Figure~\ref{1dabrc} obtained
for a Monte Carlo run on a one-dimensional lattice (no site occupation 
restrictions), indicating localized population explosion and extinction events.

\section{Correlation functions and field theory}
\label{sec:corft}

The above example demonstrates that stochastic fluctuations and correlations 
induced by the dynamical interactions may lead to important features that are 
not adequately described by mean-field approaches.
We thus require tools that allow us to systematically account for fluctuations 
in the mathematical description of stochastic complex systems and evaluate
characteristic correlations. 
Such a toolbox is provided through {\em field theory} representations that are
conducive to the identification of underlying symmetries and have proven useful
starting points for the construction of various approximation schemes.
These methods were originally devised and elaborated in the theory of (quantum
and classical) many-particle systems and quantum fields 
(Refs.~\cite{Ramond}--\cite{Cardy} represent a sample of recent textbooks).

\subsection{Generating functions}
\label{sus:genfc}

The basic structure of these field theories rests in a (normalized) exponential
probability distribution ${\cal P}[S_i]$ for the $N$ relevant variables $S_i$, 
$i=1,\ldots,N$: $\int \prod_{i=1}^N dS_i \, {\cal P}[S_i] = 1$, where the 
integration extends over the allowed range of values for the $S_i$; i.e.,
\begin{equation}
  {\cal P}[S_i] = \frac{1}{\cal Z} \ \exp\Bigl( - {\cal A}[S_i] \Bigr)\ , \quad
  {\cal Z} = \int \prod_{i=1}^N dS_i \ \exp\Bigl( - {\cal A}[S_i] \Bigr) \ .
\label{ftrepr}
\end{equation}
In canonical equilibrium statistical mechanics, 
${\cal A}[S_i] = {\cal H}[S_i] / k_{\rm B} T$ is essentially the Hamiltonian, 
and the normalization is the partition function ${\cal Z}$.
In Euclidean quantum field theory, the action ${\cal A}[S_i]$ is given by the 
Langrangian.

All observables ${\cal O}$ should be functions of the basic degrees of freedom 
$S_i$; their ensemble average thus becomes
\begin{equation}
  \Big\langle {\cal O}[S_i] \Big\rangle = \int \prod_{i=1}^N dS_i \, 
  {\cal O}[S_i] \, {\cal P}[S_i] = \frac{1}{\cal Z} \ \int \prod_{i=1}^N dS_i \
  {\cal O}[S_i] \ \exp\Bigl( - {\cal A}[S_i] \Bigr) \ .
\label{obsexp}
\end{equation}
If we are interested in $n$-point correlations, i.e., expectation values of the
products of the variables $S_i$, it is useful to define a {\em generating 
function}
\begin{equation} 
  {\cal W}[j_i] = \Big\langle \exp \sum_{i=1}^N j_i \, S_i \Big\rangle \ ,
\label{genfun}
\end{equation}
with ${\cal W}[j_i = 0] = 1$.
Notice that ${\cal W}[j_i]$ formally is just the Laplace transform of the
probability distribution ${\cal P}[S_i]$.
The correlation functions can now be obtained via partial derivatives of 
${\cal W}[j_i]$ with respect to the sources $j_i$:
\begin{equation}
  \Big\langle S_{i_1} \ldots \, S_{i_n} \Big\rangle = 
  \frac{\partial}{\partial j_{i_1}} \ldots \frac{\partial}{\partial j_{i_n}} \,
  {\cal W}[j_i] \ \Big\vert_{j_i=0} \ .
\label{corfun}
\end{equation}
Connected correlation functions or cumulants can be found by similar partial
derivatives of the logarithm of the generating function:
\begin{equation}
  \Big\langle S_{i_1} \ldots \, S_{i_n} \Big\rangle_c = 
  \frac{\partial}{\partial j_{i_1}} \ldots \frac{\partial}{\partial j_{i_n}} \,
  \ln {\cal W}[j_i] \ \Big\vert_{j_i=0} \ ,
\label{cumfun}
\end{equation}
e.g., $\langle S_i \rangle_c = \langle S_i \rangle$, and 
$\langle S_i \, S_j \rangle_c = \langle S_i \, S_j \rangle - 
\langle S_i \rangle \, \langle S_j \rangle = 
\langle (S_i - \langle S_i \rangle) \, (S_j - \langle S_j \rangle) \rangle$.

\subsection{Perturbation expansion}
\label{sus:perex}

For a Gaussian action, i.e., a quadratic form 
${\cal A}_0[S_i] = \frac12 \sum_{ij} S_i \, A_{ij} \, S_j$ (for simplicity we 
assume real variables $S_i$), one may readily compute the corresponding 
generating function ${\cal W}_0[j_i]$.
After diagonalizing the symmetric $N \times N$ matrix $A_{ij}$, completing the 
squares, and evaluating the ensuing Gaussian integrals, one obtains
\begin{equation}
  {\cal Z}_0 = \frac{(2 \pi)^{N/2}}{\sqrt{\det A}} \ , \quad 
  {\cal W}_0[j_i] = \exp \biggl( \frac12 \sum_{i,j=1}^N j_i \, A^{-1}_{ij} \, 
  j_j \biggr) , \quad \left\langle S_i \, S_j \right\rangle_0 = A^{-1}_{ij} \ .
\label{gaugen}
\end{equation}
Thus, the two-point correlation functions in the Gaussian ensemble are given by
the elements of the inverse harmonic coupling matrix.
An important special property of the Gaussian ensemble is that all $n$-point
functions with odd $n$ vanish, whereas those with even $n$ factorize into sums
of all possible permutations of products of two-point functions $A^{-1}_{ij}$ 
that can be constructed by pairing up the variables $S_i$ (Wick's theorem).
For example, the four-point function reads 
$\langle S_i \, S_j \, S_k \, S_l \rangle_0 = A^{-1}_{ij} \, A^{-1}_{kl} +
A^{-1}_{ik} \, A^{-1}_{jl} + A^{-1}_{il} \, A^{-1}_{jk}$.

Let us now consider a general action, isolate the Gaussian contribution, and 
label the remainder as the nonlinear, anharmonic, or interacting part, 
${\cal A}[S_i] = {\cal A}_0[S_i] + {\cal A}_{\rm int}[S_i]$.
We then observe that
\begin{equation}
  {\cal Z} = {\cal Z}_0 \ \Big\langle \exp\Bigl( - {\cal A}_{\rm int}[S_i]
  \Bigr) \Big\rangle_0 \ , \quad \Big\langle {\cal O}[S_i] \Big\rangle = 
  \frac{\Big\langle {\cal O}[S_i] \, \exp\Bigl( - {\cal A}_{\rm int}[S_i] 
  \Bigr) \Big\rangle_0}{\Big\langle \exp\Bigl( - {\cal A}_{\rm int}[S_i] \Bigr)
  \Big\rangle_0} \ ,
\label{redact}
\end{equation}
where the index $0$ indicates that the expectation values are computed in the
Gaussian ensemble.
The nonlinear terms in Eq.~(\ref{redact}) may now be treated perturbatively by
expanding the exponentials in the numerator and denominator with respect to the
interacting part ${\cal A}_{\rm int}[S_i]$:
\begin{equation}
  \Big\langle {\cal O}[S_i] \Big\rangle = \frac{\Big\langle {\cal O}[S_i] \
   \sum_{\ell=0}^\infty \frac{1}{\ell !} \Bigl( - {\cal A}_{\rm int}[S_i] 
   \Bigr)^\ell \Big\rangle_0}{\Big\langle \sum_{\ell=0}^\infty \frac{1}{\ell !}
   \Bigl( - {\cal A}_{\rm int}[S_i] \Bigr)^\ell \Big\rangle_0} \ .
\label{pertex}
\end{equation}
If the interaction terms are polynomial in the variables $S_i$, Wick's theorem
reduces the calculation of $n$-point functions to a summation of products of 
Gaussian two-point functions.
Since the number of contributing terms grows factorially with the order $\ell$
of the perturbation expansion, graphical representations in terms of Feynman
diagrams become very useful for the classification and evaluation of the
different contributions to the perturbation series.
Basically, they consist of lines representing the Gaussian two-point functions
(`propagators') that are connected to vertices that stem from the (polynomial)
interaction terms; for details, see, e.g., Refs.~\cite{Ramond}--\cite{Cardy}.

\subsection{Continuum limit and functional integrals}
\label{sus:field}

Discrete spatial degrees of freedom are already contained in the above formal
description: for example, on a $d$-dimensional lattice with $N_d$ sites the 
index $i$ for the fields $S_i$ merely needs to entail the site labels, and the 
total number of degrees of freedom is just $N = N_d$ times the number of 
independent relevant quantities.
Upon discretizing time, these prescriptions can be extended in effectively an 
additional dimension to systems with temporal evolution.
We may at last take the {\em continuum limit} by letting $N \to \infty$, while 
the lattice constant and elementary time step tend to zero in such a manner 
that macroscopic dynamical features are preserved.
Formally, this replaces sums over lattice sites and time steps with spatial and
temporal integrations; the action ${\cal A}[S_i]$ becomes a functional of the 
fields $S_i(x,t)$; partial derivatives turn into functional derivatives; and
functional integrations $\int \prod_{i=1}^N dS_i \to \int {\cal D}[S_i]$ are to
be inserted in the previous expressions. 
For example, Eqs.~(\ref{obsexp}), (\ref{genfun}), and (\ref{cumfun}) become
\begin{eqnarray}
  &&\Big\langle {\cal O}[S_i] \Big\rangle = \frac{1}{\cal Z} \ \int 
  {\cal D}[S_i] \ {\cal O}[S_i] \ \exp\Bigl( - {\cal A}[S_i] \Bigr) \ ,
\label{obscon} \\
  &&{\cal W}[j_i] = \Big\langle \exp \int \! d^dx \int \! dt \, 
  \sum_i j_i(x,t) \, S_i(x,t) \Big\rangle \ ,
\label{gencon} \\
  &&\Big\langle \prod_{j=1}^n S_{i_j}(x_j,t_j) \Big\rangle_c = \prod_{j=1}^n
  \frac{\delta}{\delta j_{i_j}(x_j,t_j)} \ \ln {\cal W}[j_i] \ 
  \Big\vert_{j_i=0} \ .
\label{cumcon}
\end{eqnarray}
Thus we have arrived at a continuum field theory.
Nevertheless, we may follow the procedures outlined above; specifically, the
perturbation expansion expressions (\ref{redact}) and (\ref{pertex}) still 
hold, yet with arguments $S_i(x,t)$ that are now fields depending on continuous
space-time parameters.

More than thirty years ago, Janssen and De~Dominicis independently derived a 
mapping of the stochastic kinetics defined through nonlinear Langevin equations
onto a field theory action (Janssen 1976 \cite{Jan}, De~Dominicis 1976 
\cite{DeDom}; reviewed in Janssen 1979 \cite{Janssen}).
Almost simultaneously, Doi constructed a Fock space representation and 
therefrom a stochastic field theory for classical interacting particle systems 
from the master equation describing the corresponding stochastic processes 
(Doi 1976 \cite{Doia, Doib}).
His approach was further developed by several authors into a powerful method 
for the study of internal noise and correlation effects in reaction-diffusion 
systems (Grassberger and Scheunert 1980 \cite{GraSch}, Peliti 1985 
\cite{Peliti}, Peliti 1986 \cite{PelRen}, Lee 1995 \cite{Lee}, Lee and Cardy 
1995 \cite{LeeCar}; for recent reviews, see Refs.~\cite{MatGla, TauHowLee}).
We shall see below that the field-theoretic representations of both classical
master and Langevin equations require {\em two} independent fields for each 
stochastic variable.
Otherwise, the computation of correlation functions and the construction of
perturbative expansions fundamentally works precisely as sketched above.
But the underlying causal temporal structure induces important specific 
features such as the absence of `vacuum diagrams' (closed response loops): the
denominator in Eq.~(\ref{ftrepr}) is simply ${\cal Z} = 1$.
(For unified and more detailed descriptions of both versions of dynamic 
stochastic field theories, see Refs.~\cite{Tauber, UCT}.)

\section{Discrete stochastic interacting particle systems}
\label{sec:masft}

We first outline the mapping of stochastic interacting particle dynamics as 
defined through a master equation onto a field theory action 
\cite{Doia}--\cite{LeeCar}.
Let us denote the configurational probability for a stochastically evolving
system to be in state $\alpha$ at time $t$ with $P(\alpha;t)$.
Given the transition rates $W_{\alpha \to \beta}(t)$ from states $\alpha$ to 
$\beta$, a {\em master equation} essentially balances the transitions into and 
out of each state:
\begin{equation}
  \frac{\partial P(\alpha;t)}{\partial t} = 
  \sum_{\beta \not= \alpha} \Bigl[ W_{\beta \to \alpha}(t) \, P(\beta;t) 
  - W_{\alpha \to \beta}(t) \, P(\alpha;t) \,\Bigr] \ .
\label{genmas}
\end{equation}
The dynamics of many complex systems can be cast into the language of 
`chemical' reactions, wherein certain particle species (upon encounter, say)
transform into different species with fixed (time-independent) reaction rates.
The `particles' considered here could be atoms or molecules in chemistry, but
also individuals in population dynamics (as in our example in 
section~\ref{sus:lvmod}), or appropriate effective degrees of freedom governing
the system's kinetics, such as domain walls in magnets, etc.
To be specific, we envision our particles to propagate via unbiased random
walks (diffusion) on a $d$-dimensional hypercubic lattice, with the reactions
occuring according to prescribed rules when particles meet on a lattice site.
This stochastic interacting particle system is then at any time fully 
characterized by the number of particles $n_A, n_B, \ldots$ of each species 
$A, B, \ldots$ located on any lattice site.
The following describes the construction of an associated field theory action.
As important examples, we briefly discuss annihilation reactions and absorbing 
state phase transitions.

\subsection{Master equation and Fock space representation}
\label{sus:masfr}

The formal procedures are best explained by means of a simple example; thus
consider the irreversible binary annihilation process $A + A \to A$, happening
with rate $\lambda$.
In terms of the occupation numbers $n_i$ of the lattice sites $i$, we can 
construct the master equation associated with these on-site reactions as 
follows.
The annihilation process locally changes the occupation numbers by one; the
transition rate from a state with $n_i$ particles at site $i$ to $n_i - 1$
particles is $W_{n_i \to n_i - 1} = \lambda \, n_i \, (n_i - 1)$, whence
\begin{equation} 
  \frac{\partial P(n_i;t)}{\partial t} = \lambda \, (n_i + 1) \, n_i \, 
  P(n_i+1;t) - \lambda \, n_i \, (n_i-1) \, P(n_i;t)  
\label{annmas}
\end{equation} 
represents the master equation for this reaction at site $i$.
As an initial condition, we can for example choose a Poisson distribution
$P(n_i) = {\bar n}_0^{n_i} \, e^{-\bar n_0} / n_i !$ with mean initial particle
density ${\bar n}_0$.
In order to capture the complete stochastic dynamics, we just need to add 
similar contributions describing other processes, and finally sum over all 
lattice sites $i$.

Since the reactions all change the site occupation numbers by integer values,
a Fock space representation (borrowed from quantum mechanics) turns out 
particularly useful.
To this end, we introduce the harmonic oscillator or bosonic ladder operator 
algebra $[ a_i , a_j ] = 0 = [ a_i^\dagger , a_j^\dagger ]$, 
$[ a_i , a_j^\dagger ] = \delta_{ij}$, from which we construct the particle 
number eigenstates $| n_i \rangle$, namely 
$a_i \, |n_i \rangle = n_i \, |n_i-1 \rangle$, 
$a_i^\dagger \, |n_i \rangle = |n_i + 1 \rangle$,
$a_i^\dagger \, a_i \, |n_i \rangle = n_i \, |n_i \rangle$.
(Notice that a different normalization than in ordinary quantum mechanics has
been employed here.)
A general state with $n_i$ particles on sites $i$ is obtained from the `vacuum'
configuration $| 0 \rangle$, defined via $a_i \, | 0 \rangle = 0$, through the
product $| \{ n_i \} \rangle = \prod_i {a_i^\dagger}^{n_i} | 0 \rangle$. 

To implement the stochastic kinetics, we introduce a formal state vector as a
linear combination of all possible states weighted by the time-dependent 
configurational probability:
\begin{equation}
  | \Phi(t) \rangle = \sum_{\{ n_i \}} P(\{ n_i \};t) \ | \{ n_i \} \rangle \ .
\label{stavec}
\end{equation}
Simple manipulations then transform the linear time evolution according to the 
master equation into an `imaginary-time' Schr\"odinger equation
\begin{equation}   
  \frac{\partial | \Phi(t) \rangle}{\partial t} = - H \, | \Phi(t) \rangle \ ,
  \quad | \Phi(t) \rangle = e^{- H \, t} \, | \Phi(0) \rangle
\label{itschr}
\end{equation}
governed by a stochastic quasi-Hamiltonian (rather, the Liouville time 
evolution operator).
For on-site reaction processes, $H_{\rm reac} = \sum_i H_i(a_i^\dagger,a_i)$ is
a sum of local contributions; e.g., for the binary annihilation reaction,
$H_i(a_i^\dagger,a_i) = - \lambda (1 - a_i^\dagger) \, a_i^\dagger \, a_i^2$.
It is a straightforward exercise to construct the corresponding expressions 
within this formalism for the generalization $k A \to \ell A$,
\begin{equation}  
  H_i(a_i^\dagger,a_i) = - \lambda 
  \Bigl( {a_i^\dagger}^\ell - {a_i^\dagger}^k \Bigr) a_i^k \ ,
\label{reaham}
\end{equation}
and for nearest-neighbor hopping with rate $D$ between adjacent sites 
$\langle i j \rangle$,
\begin{equation}  
  H_{\rm diff} = D \sum_{<ij>} \Bigl( a_i^\dagger - a_j^\dagger \Bigr) 
  \Bigl( a_i - a_j \Bigr) \ .
\label{difham}
\end{equation} 

The two contributions for each process may be interpreted as follows: 
The first term in Eq.~(\ref{reaham}) corresponds to the actual process, and 
describes how many particles are annihilated and (re-)created in each reaction.
The second term encodes the `order' of each reaction, i.e., the number 
operator $a_i^\dagger \, a_i$ appears to the $k$th power, but in the 
normal-ordered form ${a_i^\dagger}^k \, a_i^k$, for a $k$th-order process.
These procedures are readily adjusted for reactions involving multiple particle
species.
We merely need to specify the occupation numbers on each site and 
correspondingly introduce additional ladder operators $b_i, c_i, \ldots$ for 
each new species, with $[a_i, b_i^\dagger] = 0 = [a_i, c_i^\dagger]$ etc.
For example, consider the reversible reaction 
$k A + \ell B \rightleftharpoons m C$ with forward rate $\lambda$ and backward 
rate $\sigma$; the associated reaction Hamiltonian reads
\begin{equation}  
  H_{\rm reac} = - \sum_i 
  \Bigl( {c_i^\dagger}^m - {a_i^\dagger}^k \, {b_i^\dagger}^\ell \Bigr) 
  \Bigl( \lambda \, a_i^k \, b_i^\ell - \sigma \, c_i^m \Bigr) \ .
\label{mulspe}
\end{equation}
Similarly, for the Lotka--Volterra model of section~\ref{sus:lvmod}, one finds
\begin{equation}  
  H_{\rm reac} = - \sum_i \left[ \mu \Bigl( 1 - a_i^\dagger \Bigr) a_i 
  + \sigma \Bigl( b_i^\dagger - 1 \Bigr) b_i^\dagger b_i + \lambda 
  \Bigl( a_i^\dagger - b_i^\dagger \Bigr) a_i^\dagger a_i b_i \right] \ .
\label{lovham}
\end{equation}
Note that all the above quasi-Hamiltonians are non-Hermitean operators, which 
naturally reflects the creation and destruction of particles.

Our goal is to compute averages and correlation functions with respect to the 
configurational probability $P(\{ n_i \};t)$.
Returning to a single-species system (again, the generalization to many 
particle species is obvious), this is accomplished with the aid of the 
projection state $\langle {\cal P} | = \langle 0 | \prod_i e^{a_i}$, for which 
$\langle {\cal P} | 0 \rangle = 1$ and 
$\langle {\cal P} | a_i^\dagger = \langle {\cal P} |$, since 
$[ e^{a_i} , a_j^\dagger ] = e^{a_i} \, \delta_{ij}$.
For the desired statistical averages of observables (which must all be
expressible as functions of the occupation numbers $\{ n_i \}$), one obtains 
\begin{equation}
  \langle {\cal O}(t) \rangle = \sum_{\{ n_i \}} {\cal O}(\{ n_i \}) \, 
  P(\{ n_i \};t) = \langle {\cal P} | \, {\cal O}(\{ a_i^\dagger \, a_i \}) \, 
  | \Phi(t) \rangle \ .
\label{masave}
\end{equation}
For example, as a consequence of probability conservation, 
$1 = \langle {\cal P} | \Phi(t) \rangle = \langle {\cal P} | e^{- H \, t} | 
\Phi(0) \rangle$.
Thus necessarily $\langle {\cal P} | H = 0$; upon commuting $e^{\sum_i a_i}$ 
with $H$, the creation operators are shifted $a_i^\dagger \to 1 + a_i^\dagger$,
whence this condition is fulfilled provided $H_i(a_i^\dagger \to 1,a_i) = 0$, 
which is indeed satisfied by our above explicit expressions (\ref{reaham}) and 
(\ref{difham}).
Through this prescription, we may replace $a_i^\dagger \, a_i \to a_i$ in all
averages; e.g., the particle density becomes $a(t) = \langle a_i(t) \rangle$.

In the bosonic operator representation above, we have assumed that no 
restrictions apply to the particle occupation numbers $n_i$ on each site.
If $n_i \leq 2 s + 1$, one may instead employ a representation in terms of spin
$s$ operators.
For example, particle exclusion systems with $n_i = 0$ or $1$ can thus be
mapped onto non-Hermitean spin $1/2$ `quantum' systems (for recent overviews,
see Refs.~\cite{Schuetz, Stinch}).
Specifically in one dimension, such representations in terms of integrable
spin chains have been very fruitful.
An alternative approach uses the bosonic theory, but incorporates the site
occupation restrictions through exponentials in the number operators 
$e^{- a_i^\dagger a_i}$ (van Wijland 2001 \cite{Wijland}).

\subsection{Continuum limit and field theory}
\label{sus:conft}

As a next step, we follow an established route in quantum many-particle theory 
\cite{NegOrl} and proceed towards a field theory representation through 
constructing the path integral equivalent to the `Schr\"odinger' dynamics 
(\ref{itschr}) based on coherent states, which are right eigenstates of the 
annihilation operator, 
$a_i \, |\phi_i \rangle = \phi_i \, |\phi_i \rangle$, with complex
eigenvalues $\phi_i$.
Explicitly, $|\phi_i \rangle = \exp\left( - \frac12 \, |\phi_i|^2 + \phi_i \, 
a_i^\dagger \right) | 0 \rangle$, and these coherent states satisfy the 
overlap formula $\langle \phi_j |\phi_i \rangle = \exp\left( - 
\frac12 |\phi_i|^2 - \frac12 |\phi_j|^2 + \phi_j^* \, \phi_i \right)$, and the
(over-)completeness relation $\int \prod_i d^2 \phi_i \, 
|\{ \phi_i \} \rangle \, \langle \{ \phi_i \}| = \pi$.
Upon splitting the temporal evolution (\ref{itschr}) into infinitesimal 
increments, standard procedures (elaborated in detail in Ref.~\cite{TauHowLee})
eventually yield an expression for the configurational average 
\begin{equation} 
  \langle {\cal O}(t) \rangle \propto \int \prod_i d\phi_i \, d\phi_i^* \, 
  {\cal O}(\{ \phi_i \}) \ e^{- {\cal A}[\phi_i^*,\phi_i;t]} \ ,
\label{copain}
\end{equation}
which is of the form (\ref{obsexp}), with the action
\begin{equation}
  {\cal A}[\phi_i^*,\phi_i;t_f] = \sum_i \biggl( - \phi_i(t_f) 
  + \int_0^{t_f} \! dt \left[ \phi_i^* \, \frac{\partial \phi_i}{\partial t}
  + H_i(\phi_i^*,\phi_i) \right] - {\bar n}_0 \, \phi^*_i(0) \biggr) \ ,
\label{cohact} 
\end{equation}
where the first term originates from the projection state, and the last one
stems from the initial Poisson distribution.
Through this procedure, in the original quasi-Hamiltonian the creation and 
annihilation operators $a_i^\dagger$ and $a_i$ are simply replaced with the 
complex numbers $\phi_i^*$ and $\phi_i$.

Finally, we proceed to the continuum limit, $\phi_i(t) \to \psi(\vec{x},t)$,  
$\phi_i^*(t) \to {\hat \psi}(\vec{x},t)$.
The `bulk' part of the action then becomes
\begin{equation} 
  {\cal A}[{\hat \psi},\psi] = \int \! \D^dx \! \int \! \D t \left[  
  {\hat \psi} \left( \frac{\partial}{\partial t} - D \, \vec{\nabla}^2 \right)
  \psi + {\cal H}_{\rm reac}({\hat \psi},\psi) \right] \ ,  
\label{masfth} 
\end{equation}
where the discrete hopping contribution (\ref{difham}) has naturally turned 
into a continuum diffusion term.
We have thus arrived at a {\em microscopic} field theory for stochastic
reaction--diffusion processes, without invoking any assumptions on the form or 
correlations of the internal reaction noise.
Note that we require two independent fields ${\hat \psi}$ and $\psi$ to capture
the stochastic dynamics.
Actions of the type (\ref{masfth}) may serve as a basis for further systematic 
coarse-graining, constructing a perturbation exapnsion as outlined in
section~\ref{sus:perex}, and perhaps a subsequent renormalization group 
analysis \cite{TauHowLee}--\cite{UCT}.
We remark that it is often useful to perform a shift in the field ${\hat \psi}$
about the mean-field solution, ${\hat \psi}(x,t) = 1 + {\widetilde \psi}(x,t)$.
For occasionally, the resulting field theory action allows the derivation of an
equivalent Langevin dynamics, see section~\ref{sec:lanft} below.

\subsection{Annihilation processes}
\label{sus:annpr}

Let us consider our simple single-species example $k A \to \ell A$.
The reaction part of the corresponding field theory action reads
\begin{equation}
  {\cal H}_{\rm reac}({\hat \psi},\psi) = - \lambda \left( {\hat \psi}^\ell - 
  {\hat \psi}^k \right) \psi^k\ , 
\label{annfth}
\end{equation}
see Eq.~(\ref{reaham}). 
It is instructive to study the {\em classical field equations}, namely  
$\delta {\cal A} / \delta \psi = 0$, which is always solved by 
${\hat \psi} = 1$, reflecting probability conservation, and 
$\delta {\cal A} / \delta {\hat \psi} = 0$, which, upon inserting 
${\hat \psi} = 1$ yields
\begin{equation}  
  \frac{\partial \psi(x,t)}{\partial t} = D \, \nabla^2 \, \psi(x,t) 
  - (k - \ell) \, \lambda \, \psi(x,t)^k \ ,
\label{clfleq}
\end{equation}
i.e., the mean-field equation for the local particle density $\psi(x,t)$, 
supplemented with a diffusion term.
For $k = 1$, the particle density grows ($k < \ell$) or decays ($k > \ell$) 
exponentially.
The solution of the rate equation for $k > 1$, $a(t) = \langle \psi(x,t) 
\rangle = \left[ a(0)^{1-k} + (k-l) (k-1) \, \lambda \, t \right]^{-1/(k-1)}$
implies a divergence within a finite time for $k < \ell$, and an algebraic 
decay $\sim (\lambda \, t)^{-1/(k-1)}$ for $k > \ell$.

The full field theory action, which was derived from the master equation 
defining the very stochastic process, provides a means of systematically 
including fluctuations in the mathematical treatment.
Through a dimensional analysis, we can determine the (upper) {\em critical 
dimension} below which fluctuations become sufficiently strong to alter these 
power laws.
Introducing an inverse length scale $\kappa$, $[x] \sim \kappa^{-1}$, and 
applying diffusive temporal scaling, $[D \, t] \sim \kappa^{-2}$, and
$[{\hat \psi}(x,t)] \sim \kappa^0$, $[\psi(x,t)] \sim \kappa^d$ in $d$ spatial
dimensions, the reaction rate in terms of the diffusivity scales according to
$[\lambda / D] \sim \kappa^{2 - (k-1) d}$.
In large dimensions, the kinetics is {\em reaction-limited}, and at least 
qualitatively correctly described by the mean-field rate equation.
In low dimensions, the dynamics becomes {\em diffusion-limited}, and the 
annihilation reactions generate depletion zones and spatial particle 
anti-correlations that slow down the density decay.
The nonlinear coupling $\lambda / D$ becomes dimensionless at the boundary 
critical dimension $d_c(k) = 2 / (k-1)$ that separates these two distinct 
regimes.
Thus in physical dimensions, intrinsic stochastic fluctuations are relevant
only for pair and triplet annihilation reactions.
By means of a renormalization group analysis (for details, see 
Ref.~\cite{TauHowLee}) one finds for $k = 2$ and $d < d_c(2) = 2$: 
$a(t) \sim (D \, t)^{-d/2}$ \cite{PelRen, Lee}, as confirmed by exact 
solutions in one dimension.
Precisely at the critical dimension, the mean-field decay laws acquire 
logarithmic corrections, namely $a(t) \sim (D \, t)^{-1} \ln (D \, t)$ for 
$k = 2$ at $d_c(2) = 2$, and 
$a(t) \sim \left[ (D \, t)^{-1} \ln (D \, t) \right]^{1/2}$ for $k = 3$ at 
$d_c(3) = 1$. 
Annihilation reaction between different species (e.g., $A + B \to \emptyset$) 
may introduce additional correlation effects, such as particle segregation and
the confinement of active dynamics to narrow reaction zones \cite{LeeCar}; a
recent overview can be found in Ref.~\cite{TauHowLee}.

\subsection{Active to absorbing state phase transitions}
\label{sus:abstr}

Competition between particle production and decay processes leads to even 
richer scenarios, and can induce genuine nonequilibrium transitions that 
separate `active' phases (wherein the particle densities remain nonzero in the 
long-time limit) from `inactive' stationary states (where the concentrations
ultimately vanish).
A special but abundant case are {\em absorbing states}, where, owing to the 
absence of any agents, stochastic fluctuations cease entirely, and no particles
can be regenerated \cite{ChoDro, MarDic}.
These occur in a variety of systems in nature (Refs.~\cite{Hinric, Odor} 
contain extensive discussions of various model systems); examples are chemical 
reactions involving an inert state $\emptyset$, wherefrom no reactants $A$ are
released anymore, or stochastic population dynamics models, combining diffusive
migration of a species $A$ with asexual reproduction $A \to 2 A$ (with rate 
$\sigma$), spontaneous death $A \to \emptyset$ (at rate $\mu$), and lethal 
competition $2 A \to A$ (with rate $\lambda$). 
In the inactive state, where no population members $A$ are left, clearly all 
processes terminate.
Similar effective dynamics may be used to model certain nonequilibrium physical
systems, such as the domain wall kinetics in Ising chains with competing 
Glauber and Kawasaki dynamics. 
Here, spin flips $\uparrow \uparrow \downarrow \downarrow \, \to \,
\uparrow \uparrow \uparrow \downarrow$ and $\uparrow \uparrow 
\downarrow \uparrow \, \to \, \uparrow \uparrow \uparrow \uparrow$ 
may be viewed as domain wall ($A$) hopping and pair annihilation 
$2 A \to \emptyset$, whereas spin exchange $\uparrow \uparrow \downarrow 
\downarrow \, \to \, \uparrow \downarrow \uparrow \downarrow$ represents a 
branching process $A \to 3 A$. 
Notice that the para- and ferromagnetic phases respectively map onto the active
and inactive `particle' states.
The ferromagnetic state becomes absorbing if the spin flip rates are taken at 
zero temperature.

The reaction quasi-Hamiltonian corresponding to the stochastic dynamics of the
aforementioned population dynamics model reads 
\begin{equation}
  {\cal H}_{\rm reac}({\hat \psi},\psi) = \left( 1 - {\hat \psi} \right) 
  \left( \sigma \, {\hat \psi} \psi - \mu \, \psi 
  - \lambda \, {\hat \psi} \psi^2 \right) \ .
\label{fisfth}
\end{equation}
The associated rate equation is the Fisher--Kolmogorov equation (see Murray 
2002 \cite{Murray})
\begin{equation}
  \dot{a}(t) = \left( \sigma - \mu \right) a(t) - \lambda \, a(t)^2 \ ,
\label{fisher}
\end{equation}
which yields both inactive and active phases:
For $\sigma < \mu$ we have $a(t \to \infty) \to 0$, whereas for $\sigma > \mu$ 
the density eventually saturates at $a_s = (\sigma - \mu) / \lambda$.
The explicit time-dependent solution $a(t) = 
a(0) \, a_s \Big/ \Bigl[ a(0) + [a_s - a(0)] \, e^{(\mu - \sigma) t} \Bigr]$
shows that both stationary states are approached exponentially in time.
They are separated by a continuous nonequilibrium phase transition at 
$\sigma = \mu$, where the temporal decay becomes algebraic, 
$a(t) = a(0) / [1 + a(0) \lambda \, t]) \to 1 / (\lambda \, t)$ as 
$t \to \infty$, independent of the initial density $a(0)$.
As in second-order equilibrium phase transitions, however, critical 
fluctuations are expected to invalidate the mean-field power laws in low 
dimensions $d < d_c$.

If we now shift the field ${\hat \psi}$ about its stationary value $1$ and 
rescale according to ${\hat \psi}(\vec{x},t) = 1 + \sqrt{\sigma / \lambda} 
\, {\widetilde S}(\vec{x},t)$ and 
$\psi(\vec{x},t) = \sqrt{\lambda / \sigma} \, S(\vec{x},t)$, the (bulk) action 
becomes
\begin{equation}  
  {\cal A}[{\widetilde S},S] = \int \! d^dx \! \int \! dt \, \left[ 
  {\widetilde S} \left( \frac{\partial}{\partial t} + D \left( 
  r - \vec{\nabla}^2 \right) \right) S - u \left( {\widetilde S} - S \right) 
  {\widetilde S} \, S + \lambda \, {\widetilde S}^2 \, S^2 \right] \ .
\label{regfth} 
\end{equation} 
Thus, the three-point vertices have been scaled to identical coupling
strengths $u = \sqrt{\sigma \, \lambda}$, which in fact represents the 
effective coupling of the perturbation expansion.
Its scaling dimension is $[u] = \mu^{2-d/2}$, whence we infer the upper 
critical dimension $d_c = 4$.
The four-point vertex $\propto \lambda$, with $[\lambda] = \mu^{2-d}$, is then
found to be irrelevant in the renormalization group sense, and can be dropped 
for the computation of universal, asymptotic scaling properties. 
The action (\ref{regfth}) with $\lambda = 0$ is known as Reggeon field theory
(Moshe 1978 \cite{Moshe}); it satisfies a characteristic symmetry, namely 
invariance under so-called rapidity inversion 
$S(\vec{x},t) \leftrightarrow - {\widetilde S}(\vec{x},-t)$.
Remarkably, it has moreover been established that the field theory action 
(\ref{regfth}) describes the scaling properties of critical directed 
percolation clusters \cite{Obuk, CarSug, JanDP}.
The fluctuation-corrected universal power laws governing the vicinity of the
phase transition can be extracted by renormalization group methods (reviewed 
for directed percolation in Ref.~\cite{JanTau}).
Table~\ref{dpexpt} compares the analytic results obtained in an $\epsilon$ 
expansion about the critical dimension ($\epsilon = 4 - d$) with the critical 
exponent values measured in Monte Carlo computer simulations \cite{Hinric, 
Odor}.
\begin{table}  
\centering  
\caption{{\bf Field-theoretic Methods.} 
  Comparison of the values for the critical exponents of the directed 
  percolation universality class measured in Monte Carlo simulations with the 
  analytic renormalization group results within the $\epsilon = 4 - d$ 
  expansion: $\xi$ denotes the correlation length, $t_c$ the characteristic 
  relaxation time, $a_s$ the saturation density in the active state, and 
  $a_c(t)$ the critical density decay law.} 
\label{dpexpt}
\begin{tabular}{llll}  
\hline\noalign{\smallskip}
Scaling exponent & $\ d = 1$ & $\ d = 2$ & $\ d = 4 - \epsilon$ \\
\noalign{\smallskip}\hline\noalign{\smallskip}
$\xi \sim |\tau|^{- \nu}$ & $\ \nu \approx 1.100$ & $\ \nu \approx 0.735$ &  
$\ \nu = 1/2 + \epsilon / 16 + O(\epsilon^2)$ \\
$t_c \sim \xi^z \sim |\tau|^{- z \nu}$ & $\ z \approx 1.576$ & 
$\ z \approx 1.73$ & $\ z = 2 - \epsilon / 12 + O(\epsilon^2)$ \\ 
$a_s \sim |\tau|^\beta$ & $\ \beta \approx 0.2765$ & 
$\ \beta \approx 0.584$ & $\ \beta = 1 - \epsilon / 6 + O(\epsilon^2)$ \\ 
$a_c(t) \sim t^{- \alpha}$ & $\ \alpha \approx 0.160$ & 
$\ \alpha \approx 0.46$ & $\ \alpha = 1 - \epsilon / 4 + O(\epsilon^2)$ \\ 
\noalign{\smallskip}\hline
\end{tabular}
\end{table}

According to a conjecture originally formulated by Janssen and Grassberger, any
continuous nonequilibrium phase transition from an active to an absorbing state
in a system governed by Markovian stochastic dynamics that is decoupled from 
any other slow variable, and in the absence of special additional symmetries or
quenched randomness, should in fact fall in the directed percolation 
universality class (Janssen 1981 \cite{JanDP}), Grassberger 1982 
\cite{Grassb}).
This statement has indeed been confirmed in a large variety of model sytems
(many examples are listed in Refs.~\cite{Hinric, Odor}).
It even pertains to multi-species generalizations (Janssen 2001 \cite{Hannes}),
and applies for instance to the predator extinction threshold in the stochastic
Lotka--Volterra model with restricted site occupation numbers mentioned in
section~\ref{sus:lvmod} \cite{MobGeoTau}.

\section{Stochastic differential equations}
\label{sec:lanft}

This section explains how dynamics governed by {\em Langevin-type stochastic 
differential equations} can be represented through a field-theoretic formalism
\cite{Jan, DeDom, Janssen}.
Such a description is especially useful to capture the effects of external
noise on the temporal evolution of the relevant quantities under consideration,
which encompasses the case of thermal noise induced by the coupling to a heat
bath in thermal equilibrium at temperature $T$.
The underlying assumption in this approach is that there exists a natural 
{\em separation of time scales} between the slow variables $S_i$, and all other
degrees of freedom $\zeta_i$ which in comparison fluctuate rapidly, and are
therefore summarily gathered in zero-mean noise terms, assumed to be 
uncorrelated in space and time,
\begin{equation} 
  \left\langle \zeta_i(x,t) \right\rangle = 0 \ , \quad
  \left\langle \zeta_i(x,t) \, \zeta_j(x',t') 
  \right\rangle = 2 L_{ij}[S_i] \, \delta(x-x') \, \delta(t-t') \ .
\label{gennoi}
\end{equation}
Here, the noise correlator $2 L_{ij}[S_i]$ may be a function of the slow system
variables $S_i$, and also contain operators such as spatial derivatives.
A general set of coupled Langevin-type stochastic differential equations then
takes the form
\begin{equation}
  \frac{\partial S_i(t)}{\partial t} = F_i[S_i] + \zeta_i \ ,
\label{genlan}
\end{equation}
where we may decompose the `systematic forces' into reversible terms of 
microscopic origin and relaxational contributions that are induced by the noise
and drive the system towards its stationary state (see below), i.e.: 
$F_i[S_i] = F^{\rm rev}_i[S_i] + F^{\rm rel}_i[S_i]$.
Both ingredients may contain nonlinear terms as well as mode couplings between 
different variables.
Again, we first introduce the abstract formalism, and then proceed to discuss
relaxation to thermal equilibrium as well as some examples for nonequilibrium 
Langevin dynamics.

\subsection{Field theory representation of Langevin equations}
\label{sec:laneq}

The shortest and most general route towards a field theory representation of
the Langevin dynamics (\ref{genlan}) with noise correlations (\ref{gennoi})
starts with one of the most elaborate ways to expand unity, namely through a
product of functional delta functions (for the sake of compact notations, we
immediately employ a functional integration language, but in the end all the
path integrals are defined through appropriate discretizations in space and 
time):
\begin{eqnarray} 
  1 &=& \int \prod_i {\cal D}[S_i] \prod_{(x,t)} \, 
  \delta\left( \frac{\partial S_i(x,t)}{\partial t} - F_i[S_i](x,t) 
  - \zeta_i(x,t) \right)
\label{delidn} \\ 
  &=& \int \prod_i {\cal D}[i {\widetilde S}_i] \, {\cal D}[S_i] \exp \biggl[ 
  - \int \! d^dx \! \int \! dt \sum_i {\widetilde S}_i \left( 
  \frac{\partial S_i}{\partial t} - F_i[S_i] - \zeta_i \right) \biggr] \ .
  \nonumber
\end{eqnarray}  
In the second line we have used the Fourier representation of the (functional) 
delta distribution by means of the purely imaginary auxiliary variables 
${\widetilde S}_i$ (also called Martin--Siggia--Rose response fields
\cite{MaSiRo}).
Next we require the explicit form of the noise probability distribution that
generates the correlations (\ref{gennoi}); for simplicity, we may employ the
Gaussian
\begin{equation} 
  {\cal W}[\zeta_i] \propto \exp \biggl[ - \frac14 \int \! d^dx \! \int_0^{t_f}
  \! dt \sum_{ij} \zeta_i(x,t) \left[ L_{ij}^{-1} \, \zeta_j(x,t) \right] 
  \biggr] \ . 
\label{progau}
\end{equation} 

Inserting the identity (\ref{delidn}) and the probability distribution 
(\ref{progau}) into the desired stochastic noise average of any observable 
${\cal O}[S_i]$, we arrive at  
\begin{eqnarray}
  \langle {\cal O}[S_i] \rangle_\zeta &\propto& \int \prod_i 
  {\cal D}[i {\widetilde S}_i] \, {\cal D}[S_i] \exp \biggl[ - \int \! d^dx \! 
  \int \! dt \sum_i {\widetilde S}_i \left(\frac{\partial S_i}{\partial t} 
  - F_i[S_i] \right) \biggr] \, {\cal O}[S_i] \nonumber \\ 
  &\times& \int \prod_i {\cal D}[\zeta_i] \exp \biggl( - \int \! d^dx \! 
  \int \! dt \sum_i \biggl[ \, \frac14 \, \zeta_i \sum_j L_{ij}^{-1} \,
  \zeta_j - {\widetilde S}_i \, \zeta_i \biggr] \biggr) \ .
\label{nouave}
\end{eqnarray} 
Subsequently evaluating the Gaussian integrals over the noise $\zeta_i$ yields
at last 
\begin{equation}
  \langle {\cal O}[S_i] \rangle_\zeta = \int \prod_i {\cal D}[S_i] \, 
  {\cal O}[S_i] \, {\cal P}[S_i] \ , \quad 
  {\cal P}[S_i] \propto \int \prod_i {\cal D}[i {\widetilde S}_i] \,  
  e^{- {\cal A}[{\widetilde S}_i,S_i]} \ ,
\label{noipro}
\end{equation} 
with the statistical weight governed by the Janssen--De~Dominicis `response' 
functional \cite{Jan, DeDom} 
\begin{equation}  
  {\cal A}[{\widetilde S}_i,S_i] = \int \! d^dx \! \int_0^{t_f} \! dt \sum_i 
  \left[ {\widetilde S}_i \left( \frac{\partial S_i}{\partial t} - F_i[S] 
  \right) - {\widetilde S}_i \sum_j L_{ij} \, {\widetilde S}_j \right] \ . 
\label{janded} 
\end{equation}
It should be noted that in the above manipulations, we have omitted the 
functional determinant from the variable change $\{ \zeta_i \} \to \{ S_i \}$.
This step can be justified through applying a forward (It\^o) discretization 
(for technical details, see Refs.~\cite{BaJaWa, Janssen, UCT}).
Normalization implies $\int \prod_i {\cal D}[i {\widetilde S}_i] \, 
{\cal D}[S_i] \, e^{- {\cal A}[{\widetilde S}_i,S]_i} = 1$.
The first term in the action (\ref{janded}) encodes the temporal evolution
according to the systematic terms in the Langevin equations (\ref{genlan}),
whereas the second term specifies the noise correlations (\ref{gennoi}).
Since the auxiliary fields appear only quadratically, they could be eliminated 
via completing the squares and Gaussian integrations.
This results in the equivalent Onsager--Machlup functional which however
contains squares of the nonlinear terms and the inverse of the noise
correlator operators; the form (\ref{janded}) is therefore usually more 
convenient for practical purposes.
The Janssen--De~Dominicis functional (\ref{janded}) takes the form of a
($d+1$)-dimensional statistical field theory with again {\em two} independent 
sets of fields $S_i$ and ${\widetilde S}_i$.
It may serve as a starting point for systematic approximation schemes including
perturbative expansions, and subsequent renormalization group treatments.
Causality is properly incorporated in this formalism which has important
technical implications \cite{BaJaWa, Janssen, UCT}.

\subsection{Thermal equilibrium and relaxational critical dynamics}
\label{sec:eqrel}

Consider the dynamics of a system that following some external perturbation 
relaxes towards thermal equilibrium governed by the canonical Boltzmann 
distribution at fixed temperature $T$,
\begin{equation} 
  {\cal P}_{\rm eq}[S_i] = \frac{1}{{\cal Z}(T)} \ 
  \exp\left( -{\cal H}[S_i] / k_{\rm B}T \right) \ .
\label{boldis}
\end{equation} 
The relaxational term in the Langevin equation (\ref{genlan}) can then be
specified as
\begin{equation}
  F^{\rm rel}_i[S_i] = - \lambda_i \, \frac{\delta {\cal H}[S_i]}{\delta S_i} 
  \ ,
\label{reldyn}
\end{equation}
with Onsager coefficients $\lambda_i$; for nonconserved fields, $\lambda_i$ is 
a positive relaxation rate. 
On the other hand, if the variable $S_i$ is a conserved quantity (such as the
energy density), there is an associated continuity equation 
$\partial S_i / \partial t + \nabla \cdot J_i = 0$, with a conserved current
that is typically given by a gradient of the field $S_i$: 
$J_i = - D_i \, \nabla S_i + \ldots$; as a consequence, the fluctuations of the
fields $S_i$ will relax diffusively with diffusivity $D_i$, and 
$\lambda_i = - D_i \, \nabla^2$ becomes a spatial Laplacian.

In order for ${\cal P}(t) \to {\cal P}_{\rm eq}$ as $t \to \infty$, the 
stochastic Langevin dynamics needs to satisfy {\em two} conditions, which can
be inferred from the associated Fokker--Planck equation \cite{ChaLub, UCT}.
First, the reversible probability current is required to be divergence-free in 
the space spanned by the fields $S_i$: 
\begin{equation} 
  \int \! d^dx \sum_i \frac{\delta}{\delta S_i(x)} \left( F^{\rm rev}_i[S_i] \ 
  e^{- {\cal H}[S_i] / k_{\rm B} T} \right) = 0 \ .  
\label{divcon} 
\end{equation}
This condition severely constrains the reversible force terms. 
For example, for a system whose microscopic time evolution is determined 
through the Poisson brackets 
$Q_{ij}(x,x') = \left\{ S_i(x),S_j(x') \right\} = - Q_{ji}(x',x)$ (to be 
replaced by commutators in quantum mechanics), one finds for the reversible
mode-coupling terms \cite{ChaLub}
\begin{equation}
  F^{\rm rev}_i[S_i](x) = - \int \! d^dx' \sum_j \left[ Q_{ij}(x,x') \, 
  \frac{\delta {\cal H}[S_i]}{\delta S_j(x')} - k_{\rm B} T \, 
  \frac{\delta Q_{ij}(x,x')}{\delta S_j(x')} \right] \ .
\label{modcpl}
\end{equation} 
Second, the noise correlator in Eq.~(\ref{gennoi}) must be related to the
Onsager relaxation coefficients through the Einstein relation
\begin{equation}
  L_{ij} = k_{\rm B} T \, \lambda_i \, \delta_{ij} \ . 
\label{einrel}
\end{equation}

To provide a specific example, we focus on the case of purely relaxational 
dynamics (i.e., reversible force terms are absent entirely), with the 
(mesoscopic) Hamiltonian given by the Ginzburg--Landau--Wilson free energy that
describes second-order phase transitions in thermal equilibrium for an 
$n$-component order parameter $S_i$, $i=1,\ldots,N$ \cite{Ramond}-\cite{Cardy}:
\begin{equation} 
  {\cal H}[S_i] = \int \! d^dx \sum_{i=1}^N \biggl[ \frac{r}{2} \, [S_i(x)]^2
  + \frac12 \, [\nabla S_i(x)]^2 + \frac{u}{4 !} \, [S_i(x)]^2 
  \sum_{j=1}^N [S_j(x)]^2 \biggr] \ , 
\label{glwham} 
\end{equation} 
where the control parameter $r \propto T-T_c$ changes sign at the critical
temperature $T_c$, and the positive constant $u$ governs the strength of the 
nonlinearity.
If we assume that the order parameter itself is not conserved under the
dynamics, the associated response functional reads
\begin{equation} 
  {\cal A}[{\widetilde S}_i,S_i] = \int \! d^dx \! \int \! dt \sum_i  
  {\widetilde S}_i \left( \frac{\partial}{\partial t} + \lambda_i \, 
  \frac{\delta {\cal H}[S_i]}{\delta S_i} - k_{\rm B} T \, \lambda_i \,
  {\widetilde S}_i \right) \ .
\label{modela}
\end{equation}
This case is frequently referred to as model A critical dynamics \cite{HohHal}.
For a diffusively relaxing conserved field, termed model B in the 
classification of Ref.~\cite{HohHal}, one has instead
\begin{equation} 
  {\cal A}[{\widetilde S}_i,S_i] = \int \! d^dx \! \int \! dt \sum_i  
  {\widetilde S}_i \left( \frac{\partial}{\partial t} - D_i \, \nabla^2 \, 
  \frac{\delta {\cal H}[S_i]}{\delta S_i} + k_{\rm B} T \, D_i \, \nabla^2 \, 
  {\widetilde S}_i \right) \ .
\label{modelb}
\end{equation}
Consider now the external fields $h_i$ that are thermodynamically conjugate to 
the mesoscopic variables $S_i$, i.e., 
${\cal H}(h_i) = {\cal H}(h_i = 0) - \int \! d^dx \sum_i h_i(x) \, S_i(x)$.
For the simple relaxational models (\ref{modela}) and (\ref{modelb}), we may 
thus immediately relate the dynamic susceptibility to two-point correlation 
functions that involve the auxiliary fields ${\widetilde S}_i$ \cite{BaJaWa}, 
namely
\begin{equation}
  \chi_{ij}(x-x',t-t') = 
  \frac{\delta \langle S_i(x,t) \rangle}{\delta h_j(x',t')} \bigg\vert_{h_i=0} 
  = k_{\rm B} T \, \lambda_i \left\langle S_i(x,t) \, {\widetilde S}_j(x',t') 
  \right\rangle  
\label{dysusa}
\end{equation}
for nonconserved fields, while for model B dynamics 
\begin{equation}
  \chi_{ij}(x-x',t-t') = - k_{\rm B} T \, D_i \left\langle S_i(x,t) \, \nabla^2
  \, {\widetilde S}_j(x',t') \right\rangle \ .
\label{dysusb}
\end{equation}
Finally, in thermal equilibrium the dynamic response and correlation functions 
are related through the fluctuation--dissipation theorem \cite{BaJaWa}
\begin{equation} 
  \chi_{ij}(x-x',t-t') = \Theta(t-t') \, \frac{\partial}{\partial t'} 
  \left\langle S_i(x,t) \, S_j(x',t') \right\rangle \ .
\label{lanfdt}
\end{equation}

\subsection{Driven diffusive systems and interface growth}
\label{sec:ddsif}

We close this section by listing a few intriguing examples for Langevin sytems
that describe genuine out-of-equilibrium dynamics.
First, consider a driven diffusive lattice gas (an overview is provided in
Ref.~\cite{SchZia}), namely a particle system with conserved total density
with biased diffusion in a specified (`$\parallel$') direction.
The coarse-grained Langevin equation for the scalar density fluctuations thus
becomes spatially anisotropic \cite{JanSch, LeuCar},
\begin{equation}
  \frac{\partial S(x,t)}{\partial t} = D \left( \nabla_\perp^2 
  + c \, \nabla_\parallel^2 \right) S(x,t) + \frac{D \, g}{2} \, 
  \nabla_\parallel S(x,t)^2 + \zeta(x,t) \ ,
\label{ddslan}
\end{equation}
and similarly for the conserved noise with $\langle \zeta \rangle = 0$,
\begin{equation}
  \left\langle \zeta(x,t) \, \zeta(x',t') \right\rangle = 
  - 2 D \left( \nabla_\perp^2 + {\tilde c} \, \nabla_\parallel^2 \right)
  \delta(x-x') \, \delta(t-t') \ .
\label{ddsnoi}
\end{equation}
Notice that the drive term $\propto g$ breaks both the system's spatial 
reflection symmetry as well as the Ising symmetry $S \to - S$.
In one dimension, Eq.~(\ref{ddslan}) coincides with the noisy Burgers equation
\cite{FoNeSt}, and since in this case (only) the condition (\ref{divcon}) is 
satisfied, effectively represents a system with equilibrium dynamics.
The corresponding Janssen--De~Dominicis response functional reads  
\begin{equation} 
  {\cal A}[{\widetilde S},S] = \!\! \int \!\! d^dx \!\! \int \!\! dt \, 
  {\widetilde S} \Biggl[ \frac{\partial S}{\partial t} 
  - D \left( \nabla_\perp^2 + c \nabla_\parallel^2 \right) S 
  + D \left( \nabla_\perp^2 + {\tilde c} \nabla_\parallel^2 \right) 
  {\widetilde S} - \frac{D \, g}{2} \nabla_\parallel S^2 \Biggr] \, .
\label{ddsact}
\end{equation} 
It describes a `massless' theory, hence we expect the system to generically 
display scale-invariant features, without the need to tune to a special point
in parameter space.
The large-scale scaling properties can be analyzed by means of the dynamic
renormalization group \cite{JanSch, LeuCar}.

Another famous example for generic scale invariance emerging in a 
nonequilibrium system is curvature-driven interface growth, as captured by the
Kardar--Parisi--Zhang equation \cite{KaPaZh}
\begin{equation} 
  \frac{\partial S(x,t)}{\partial t} = D \, \nabla^2 S(x,t)  
  + \frac{D g}{2} \, [\nabla S(x,t)]^2 + \zeta(x,t) \ ,
\label{kpzeqn}
\end{equation}
with again $\langle \zeta \rangle = 0$ and the noise correlations
\begin{equation} 
  \left\langle \zeta(x,t) \, \zeta(x',t') \right\rangle = 2 D \, \delta(x-x') 
\delta(t-t') \ . 
\label{kpznoi}
\end{equation} 
(For more details and intriguing variants, see e.g. 
Refs.~\cite{BarSta}-\cite{Krug}.)
The associated field theory action 
\begin{equation}
  {\cal A}[{\widetilde S},S] = \int \! d^dx \int \! dt \left[ {\widetilde S} 
  \left( \frac{\partial S}{\partial t} - D \, \nabla^2 S - \frac{D g}{2} \, 
  [\nabla S]^2 \right) - D \, {\widetilde S}^2 \right]
\label{kpzact}
\end{equation}
encodes surprisingly rich behavior including a kinetic roughening transition
separating two distinct scaling regimes in dimensions $d > 2$ 
\cite{BarSta}-\cite{Krug}.

\section{Future directions}
\label{sec:direc}

The rich phenomenology in many complex systems is only inadequately captured 
within widely used mean-field approximations, wherein both statistical 
fluctuations and correlations induced by the subunits' interactions or the
system's kinetics are neglected.
Modern computational techniques, empowered by recent vast improvements in data 
storage and tact frequencies, as well as the development of clever algorithms, 
are clearly invaluable in the theoretical study of model systems displaying the
hallmark features of complexity.
Yet in order to gain a deeper understanding and to maintain control over the 
typically rather large parameter space, numerical investigations need to be
supplemented by analytical approaches.
The field-theoretic methods described in this article represent a powerful set
of tools to systematically include fluctuations and correlations in the
mathematical description of complex stochastic dynamical systems composed of 
many interacting degrees of freedom.
They have already been very fruitful in studying the intriguing physics of 
highly correlated and strongly fluctuating many-particle systems.
Aside from many important quantitative results, they have provided the basis 
for our fundamental understanding of the emergence of universal macroscopic 
features.

At the time of writing, the transfer of field-theoretic methods to problems in 
chemistry, biology, and other fields such as sociology has certainly been 
initiated, but is still limited to rather few and isolated case studies.
This is understandable, since becoming acquainted with the intricate 
technicalities of the field theory formalism requires considerable effort. 
Also, whereas it is straightforward to write down the actions corresponding the
stochastic processes defined via microscopic classical discrete master or 
mesoscopic Langevin equations, it is usually not that easy to properly extract 
the desired information about large-scale structures and long-time asymptotics.
Yet if successful, one tends to gain insights that are not accessible by any
other means.
I therefore anticipate that the now well-developed methods of quantum and
statistical field theory, with their extensions to stochastic dynamics, will 
find ample successful applications in many different areas of complexity 
science.
Naturally, further approximation schemes and other methods tailored to the
questions at hand will have to be developed, and novel concepts be devised.
I look forward to learning about and hopefully also participating in these 
exciting future developments.

\subsection*{Acknowledgements}

The author would like to acknowledge financial support through the 
U.S. National Science Foundation grant NSF DMR-0308548. This article
is dedicated to the victims of the terrible events at Virginia Tech 
on April 16, 2007.

\end{document}